\documentclass[12pt,a4paper,dvips]{article}
\usepackage{epsfig,wrapfig} 
\renewcommand{\section}[1]{\vspace{6pt} \noindent\mbox{#1} \newline \noindent}
\renewcommand{\subsection}[1]{\vspace{6pt} \noindent\mbox{\underline{#1}} 
\newline \noindent}
\renewcommand{\subsubsection}[1]{\vspace{6pt} \noindent\mbox{\underline{#1}}
\noindent}

\newfont{\sansb}{cmssbx10}
\newfont{\sans}{cmss10}

\begin{document}
{\small 25 Int. Cosmic Ray Conf., Durban 1997, Highlight Session 
\hspace{4cm} ADP-AT-97-9 \vspace{-24pt}\\}     
{\center \LARGE VERY HIGH ENERGY GAMMA RAYS FROM MARKARIAN 501
\vspace{6pt}\\} 
R.J. Protheroe$^1$, C.L. Bhat$^2$, 
P. Fleury$^3$, E. Lorenz$^4$, M. Teshima$^5$, T.C. Weekes$^6$
\vspace{6pt}\\ 
{\it $^1$Dept. of Physics and Math. Phys.,
University of Adelaide, Adelaide, SA 5005, Australia.\\ 
$^2$Bhabha Atomic Research Centre, Mumbai - 400 085, India (on
behalf of the GRACE collab.)\\ 
$^3$LPNHE - Ecole Polytechnique, 91128 Palaiseau, France (on
behalf of CAT Collab.)\\ 
$^4$Max-Planck Inst.
f\"{u}r Physik, Munich, Germany (on behalf of HEGRA Collab.)\\
$^5$Inst. for Cosmic Ray Research,
University of Tokyo, Japan (on behalf of TA Collab.)\\ 
$^6$Harvard-Smithsonian CfA, Box 97, Amado, AZ 85645,
USA (on behalf of Whipple Collab.) \vspace{-12pt}\\}

{\center ABSTRACT\\} During remarkable flaring activity in 1997
Markarian 501 was the brightest source in the sky at TeV
energies, outshining the Crab Nebula by a factor of up to 10.
Periods of flaring activity each lasting a few days were observed
simultaneously by several gamma ray telescopes.  Contemporaneous
multiwavelength observations in April 1997 show that a
substantial fraction of the total AGN power is at TeV energies
indicating that high energy processes dominate the energetics of
this object.  Rapid variability, on time scales of less than a
day, and a flat spectrum extending up to at least 10 TeV
characterize this object.  Results of 1997 observations by 6
telescopes are summarized and some of the implications of these
results are discussed.\\

\setlength{\parindent}{1cm}
\section{INTRODUCTION}
Markarian 501 is a classical Bl Lacertae object, a
sub-classification of the Blazar class of AGN (core dominated,
flat-spectrum radio, highly optically polarized and optically
violently variable). It is the second closest known BL Lac
($z=0.034$) and like the closest, Markarian 421, it is a gamma-ray
source. Both are classified as X-ray-selected BL Lacs and show
virtually no emission lines.

It was first discovered as a TeV-emitting gamma-ray source by the
Whipple group in 1995 (Quinn et al. 1996). At that time, it had
not been detected by OSSE, COMPTEL or EGRET on the CGRO in any
observing period.  At discovery the average TeV emission level
was 8\% of the Crab Nebula and there was some evidence for daily
variability. In 1996 the emission level increased and there was
stronger evidence for variability (Quinn et al. 1997). The source
was confirmed as a TeV source by the HEGRA group (Bradbury et
al. 1997).\\

\section{OBSERVATIONS DURING 1997}
In January, 1997 the Whipple group noted that it was
brighter than usual in low elevation observations. This was
confirmed in February, 1997 and the HEGRA, CAT and TA groups were
alerted. After confirmation by HEGRA and CAT the three groups
sent out a joint notification as an IAU Circular (Breslin et
al. 1997). The nightly averages, plotted as fraction of the Crab
rate, as seen by the Whipple Collaboration in 1995, 1996 and
1997, are plotted in Figure 1. The average level from February
through June, 1997 was four times that of the Crab, an increase
by a factor of 50 on the discovery level in 1995.

\begin{figure}[htbp]
\vspace{22cm} \includegraphics{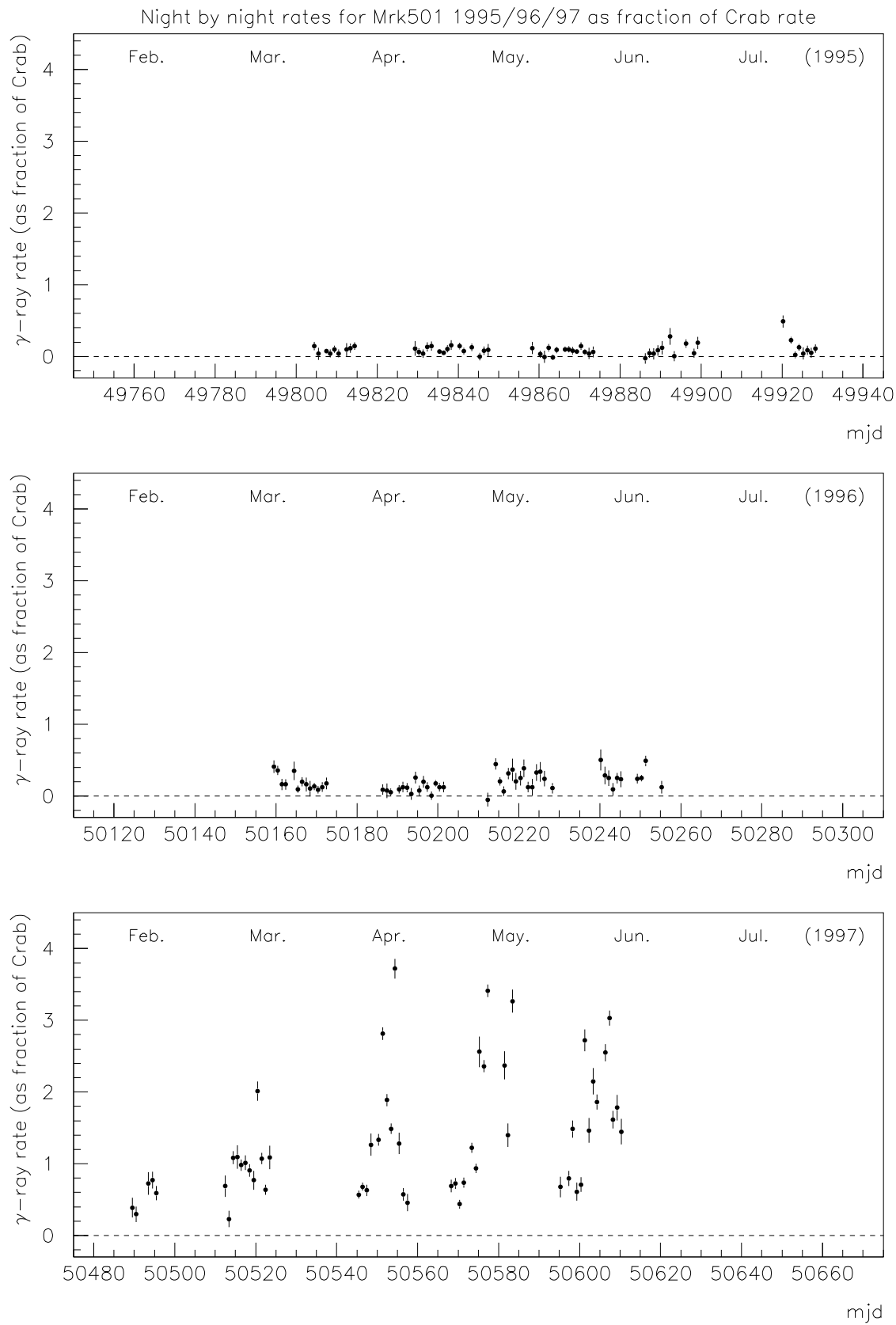}
\caption{Daily gamma-ray rates for Markarian 501 observed by Whipple
over the last 3 years expressed as fraction of average rate from
the Crab Nebula (Quinn et al. 1997)}
\end{figure}

If the 1997 Whipple observations are broken down by night into
individual 28 minute runs, there is evidence on seven nights for
significant variations using a Chi Square test. On two of these
nights the significance level of the variations is a few times
$10^{-5}$ after allowing for the number of trials (Quinn,
1997). The doubling time is approximately 2 hours.

Figure 2a shows the light curve obtained with HEGRA CT-System.
Normally, optical Cherenkov observations are not carried out if
the Moon is visible as this increases dramatically the night sky
brightness.  However, because of the strength of the gamma-ray
signal from Markarian 501 the HEGRA group continued observing
during moonlight using CT1 but with a higher energy threshold.
The complete light curve of Markarian 501 between 8 March 1997
and 9 September 1997 as observed by HEGRA CT1 is shown in Figure
2b. The different symbols indicate under which conditions the
observations were made.  In the calculation of the flux,
differences in the calibration were taken into account.

\begin{figure}[htbp]
\vspace{16cm}
\includegraphics{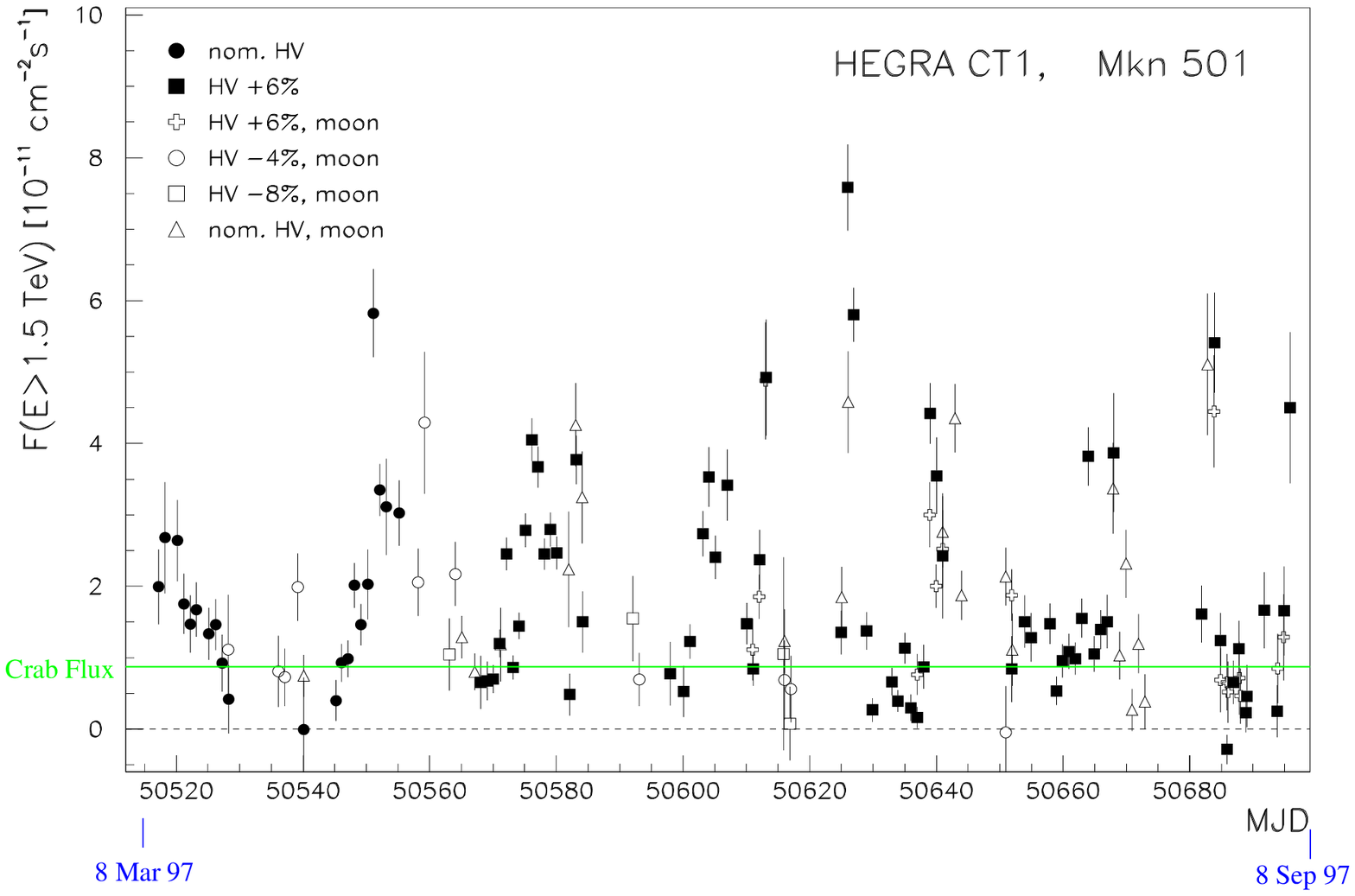}
\includegraphics{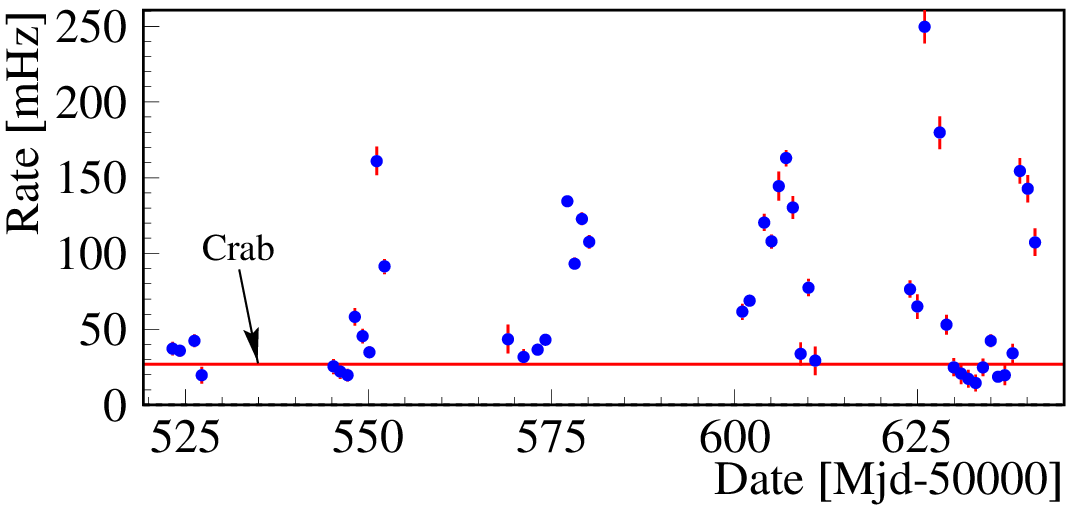}
\caption{(a) Upper panel: light curve obtained with HEGRA
    CT-System.  (b) lower panel: light curve obtained with HEGRA
    CT1. The error bars show the statistical errors. The flux
    from the Crab Nebula - as measured by the same telescope - is
    indicated by a horizontal line.}
\end{figure}

TACTIC is a compact array of 4 Cherenkov telescopes located at Mt.
Abu, India.  During its first observing campaign in April-May,
TACTIC observed Markarian-501 with a statistical significance of
14.2$\sigma$.  The corresponding source light curve is displayed
in Figure~3a.

\begin{figure}[htbp]
\vspace{23cm}
\includegraphics{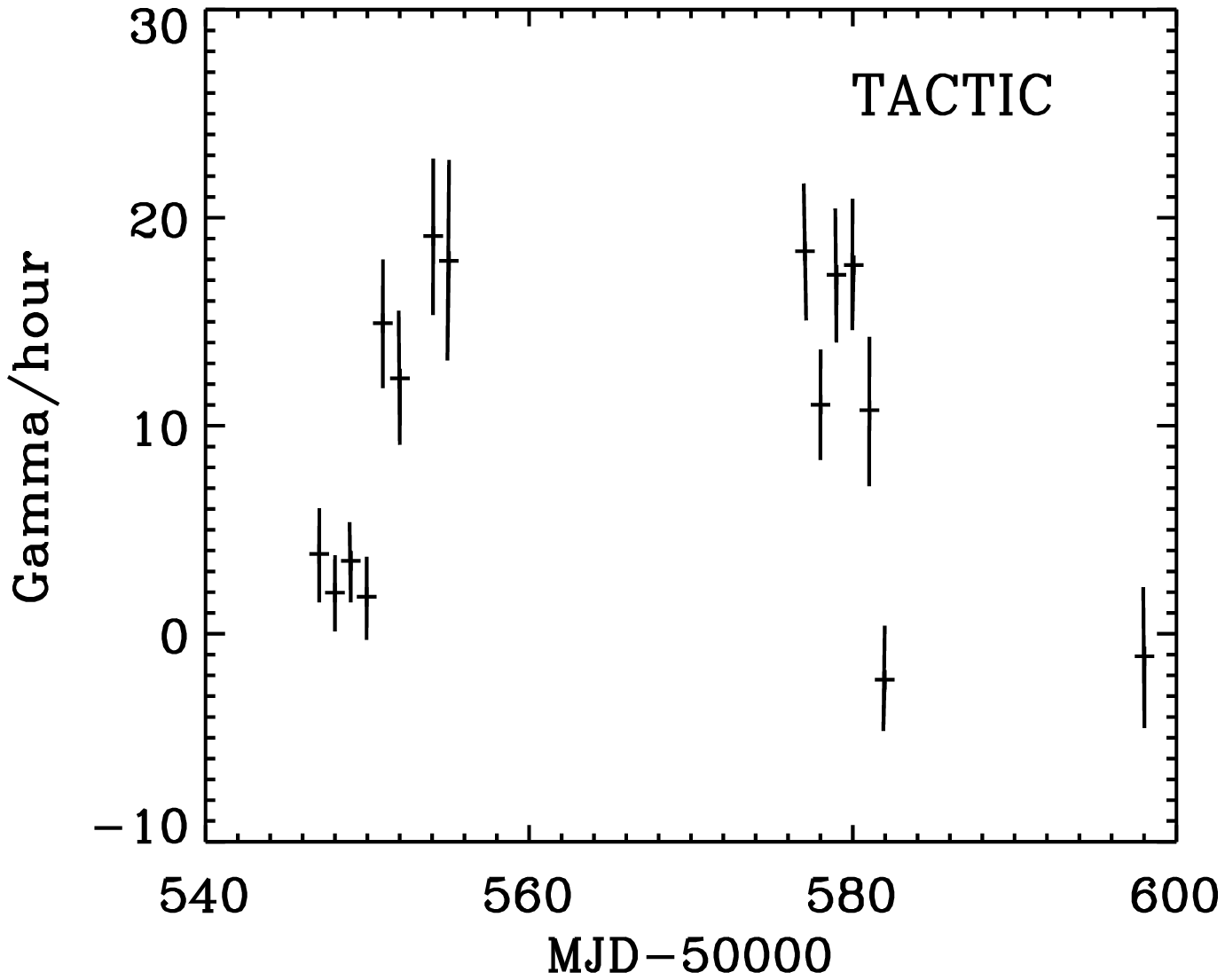}
\includegraphics{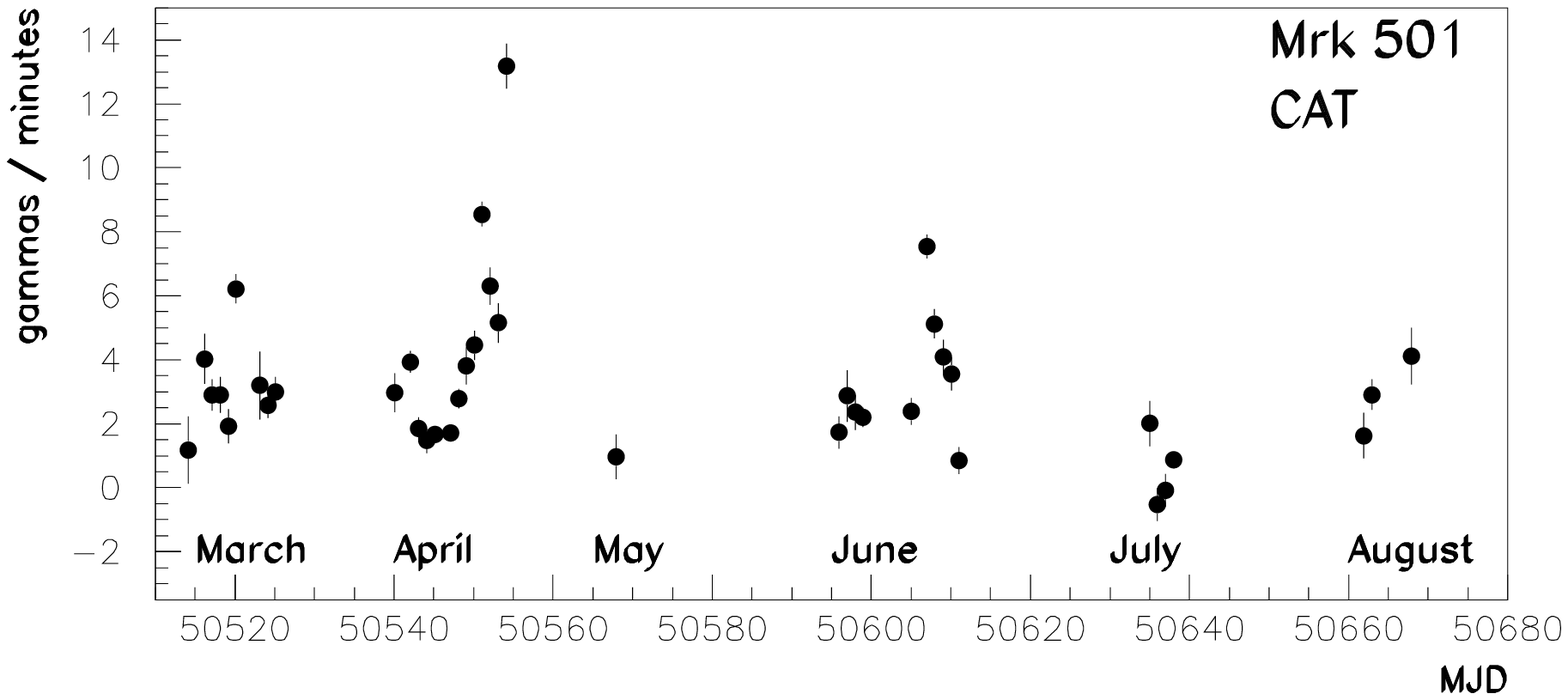}
\includegraphics{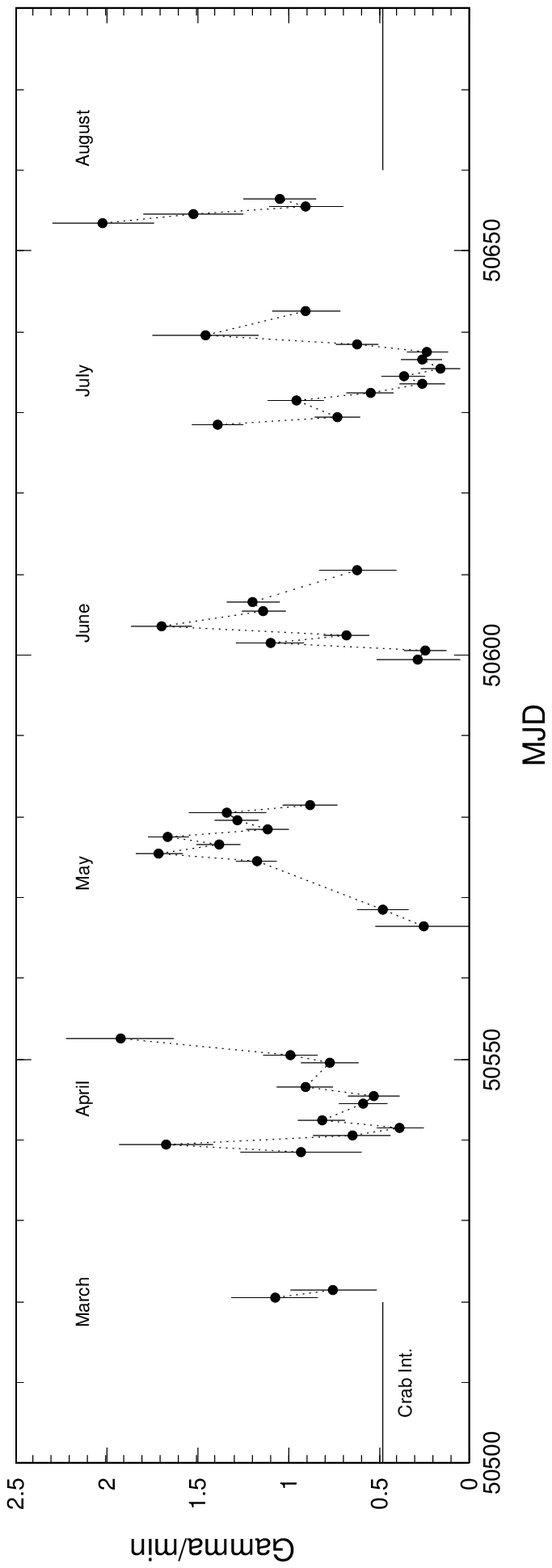}
\caption{Comparison of light curves from: (a) top -- TACTIC; (b)
middle -- CAT; (c) bottom -- TA.}
\end{figure}

CAT is an optical Cherenkov telescope located in the French
Pyrenees.  The large number of small diameter pixels in the
camera allows a relatively small mirror (5~m diameter), compared
to the Whipple telescope, to achieve a low energy threshold.  CAT
also saw Markarian 501 during its inaugural observing period and
the light curve is obtained is shown in Figure~3b.

Observations of TeV gamma ray from Markarian 501 were made using
the Telescope Array (TA) prototype located at Dugway, Utah, from
the end of March to end of July. They observed on a total number
of 47 nights. The gamma ray event rate is plotted as a function
of MJD in Figure 3c. For reference, the gamma ray rate from Crab
measured by our detectors is shown by the horizontal line.  The
event rate is highly variable day by day, and the maximum event
rate is about 5 Crab and the minimum rate is $0.5\pm0.5$ Crab.

Details of these telescopes and their observations of
Markarian 501 are given in Table~1.

\begin{center}
{\em Table 1: Telescopes used for Markarian 501 observations.\\}
\begin{tabular}{lcccccc}
\hline
\hline
\multicolumn{1}{c}{\noindent Telescope:}&
\multicolumn{1}{c}{Whipple}&
\multicolumn{2}{c}{HEGRA}&
\multicolumn{1}{c}{CAT}&
\multicolumn{1}{c}{TA}&
\multicolumn{1}{c}{TACTIC}\\
~ & ~ & CT-System & CT1 & ~ & ~ & ~ \\
\hline
Site: & Mt Hopkins & La Palma & La Palma & Th\'{e}mis & Dugway & Mt Abu \\
Longitude($^\circ$E): & -110.53 & -17.8 & -17.8 & -2.0 & -113.02 & +72.7 \\
Latitude ($^\circ$N): & 31.41 & 28.8 & 28.8 & 42.5 & 40.33 & 24.6 \\
Elevation (m): & 2300 & 2240 & 2240 & 1650 & 1600 & 1300 \\
\# telescopes: & 1 & 4 & 1 & 1 & 3 & 4 \\
Mirror area (m$^2$): & 74.0 & $4 \times 8.4$ & 5.0 & 17.5 & 
$3 \times 6.0$ & $4 \times 9.5$ \\
\# pixels: & 151 & $4 \times 271$ & 127 & 548 (+52) & 256 &81 \\
Pixel diameter ($^\circ$): & 0.25 & 0.24 & 0.24 & 0.12 & 0.25 & 0.31 \\
Threshold (GeV): & 300 & 500 & 1500 & 300 & 600 & 700 \\
Total observation: & 67.7 h & 150 h & 250 h & 88 h & 105.4 h & 50.1 h \\
February: & 3.4 h & -- & -- & 0.8 h & -- & -- \\
March: & 10.7 h & yes & yes & 16.3 h & yes & -- \\
April: & 21.4 h &  yes & yes & 39.0 h & yes & 22.54 h \\
May: & 19.8 h &  yes & yes & 7.8 h & yes & 27.51 h \\
June: & 12.4 h &  yes & yes & 11.4 h & yes & -- \\
July: & -- &  yes & yes & 5.3 h & yes  & -- \\
August: & -- & -- & yes & 5.4 h & -- & -- \\
September: & -- & -- & yes & 2.0 h & -- & -- \\
\hline
\hline
\end{tabular}
\end{center}

In addition, Markarian 501 has been detected by the University of
Durham telescope at Narrabri, Australia, by observing at
extremely large zenith angles ($\sim 72^\circ$).  The source was
also observed by the high elevation Tibet AS-$\gamma$ air
shower array which has a threshold at 10 TeV, making this the
first detection of an AGN by an air shower array.

\section{LIGHT CURVES}
All of the reported light curves (Figures 1, 2 and 3) show
excellent morphological similarity, with flaring episodes of
several days when the flux is higher than average, but also with
dramatic day to day variability.  For example, the flaring
episodes in March April, May and June were observed by all
telescopes operating at that time.  In addition to this daily
flaring, variability over periods of hours is also seen.  For
example, using the CT-System the HEGRA group investigated the
light curve on sub-hour time scales, and this is shown for the
April flare in Figure 4a.

\begin{figure}[htbp]
\vspace{13cm}
\includegraphics{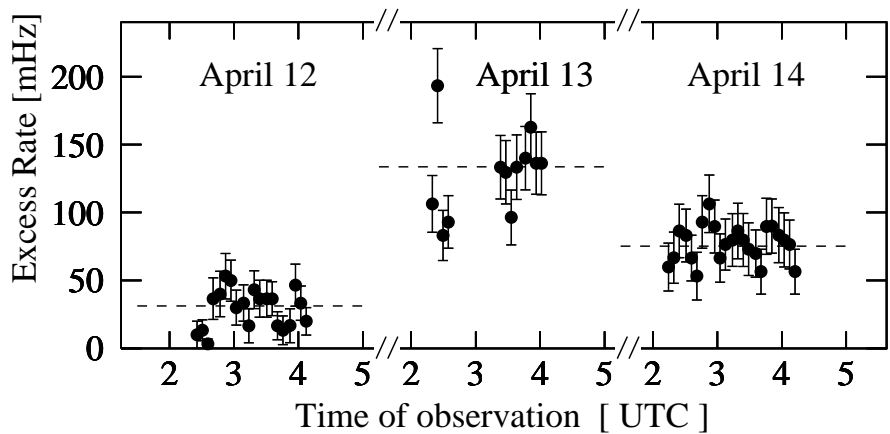}
\includegraphics{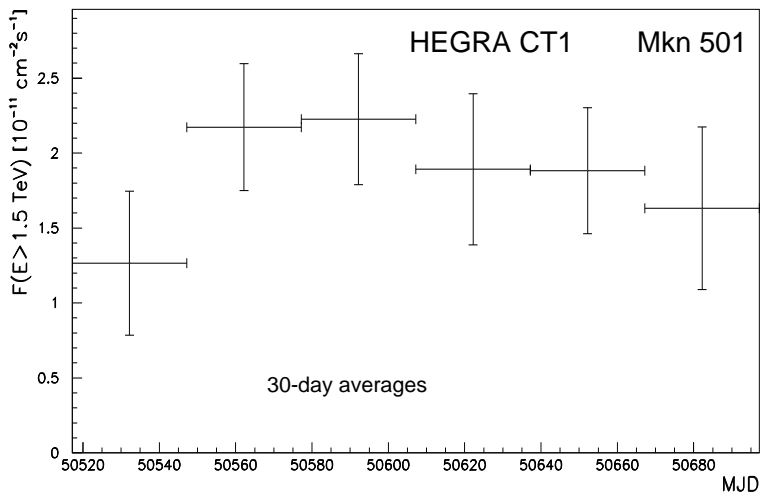}
\caption{(a) Top panel: short time scale (5 minute time bins)
behaviour during the April 1997 outburst (HEGRA CT-System).  (b)
Bottom panel: long time scale behaviour (HEGRA CT1 light curve
averaged over 30 days).}
\end{figure}

The long term behaviour (1995--1997) obtained by Whipple has been
shown in Figure 1 illustrating an increase from year to year.
Figure~4b shows the HEGRA CT1 flux averaged over periods of 30
days (due to the large daily fluctuations of the flux, the
average flux has large statistical errors). Although there is a
faint decay trend visible in the last four months of data, the
light curve averaged over 30 days is consistent with a constant
emission level during 1997.

The episodes of flaring lasting several days have occurred almost
every month, and this may indicate either a periodic or
quasiperiodic process.  For example, looking at the HEGRA CT1
light curve (Fig.~2b) we see minima every $\sim 29.6$~days.
Looking at the TA light curve (Figure 3c) we can see high states
and low states clearly the data.  This feature (the time scale
and the intensity change) in April and in July appear to be
similar, showing ``U'' shapes, and the interval between the two
high states are 14 days and 12 days.  The May and June data each
show only one high state with a ``$\Lambda$'' shape suggestive of
a possible $25.5 \pm 2$ day periodicity.  Using the Rayleigh
test, the TA group obtain a 12.7 day period with an estimated
chance probability of less than $10^{-5}$.  However, this period
requires a flare on April 13 and flaring on that day was seen by
Whipple, but both CAT and Whipple see a stronger flare on April
16.  One should also worry about possible aliasing due to gaps in
the time series (day time and full moon periods).  Thus while
very interesting, the possible periodicity should be treated as
very preliminary.\\

\section{ENERGY SPECTRUM}
Using the same analysis as was used for the Crab spectrum
(Carter-Lewis et al. 1997) and the Markarian 421 spectrum
(McEnery et al. 1997), a preliminary energy spectrum for the
Whipple observations was derived based on 255 minutes of data
(nine ON/OFF pairs) taken in April, 1997. The resulting spectrum
can be represented by:
\begin{equation}
dN/dE=(8.31 \pm 0.70) \times 10^{-7} 
\left( E \over {\rm 1 \, TeV} \right)^{-2.27 \pm 0.05} 
\; \; {\rm m^{-2} \, s^{-1} \, TeV^{-1}}.
\end{equation}
The errors shown are purely statistical. The systematic errors are
at least as large since these observations were made with an
expanded 151-pixel camera which has not yet been fully
characterized.

A series of observations at low elevations was made at Whipple to
extend the energy coverage to higher energies. These resulted in a
$5.3 \sigma$ detection in which 90\% of the gamma rays were above 8 TeV
(median energy 12 TeV) and a $3.3 \sigma$ detection in which 90\% were
above 10 TeV with a median energy of 15 TeV. It is not yet possible
to derive an energy spectrum and absolute flux with this technique.

Figure 5a shows the preliminary energy spectrum obtained by HEGRA CT1
in 1997 and is compared to that obtained in 1996. While in 1996 a
differential spectral index of $2.5\pm0.4$ was measured (open
circles), the 1997 spectrum seems to be steeper.  The excess rate
is essentially unchanged around 10 TeV. Further analysis of the
spectrum is underway.

The CAT group have obtained energy spectra during the April 15/16
flare, between flares, and for all the data (Fig.~5b).  Fits
between 200~GeV and 2~TeV for the 3 cases give differential
fluxes at 1~TeV ($10^{-7}$ m$^{-2}$ s$^{-1}$ TeV$^{-1}$) of $16
\pm 2$, $2.2 \pm 0.3$, and $5.0 \pm 0.5$ respectively.  The
corresponding spectral indices were $2.27 \pm 0.11$, $2.65 \pm
0.16$ and $2.33 \pm 0.09$.

\begin{figure}[htbp]
\vspace{22cm}
\includegraphics{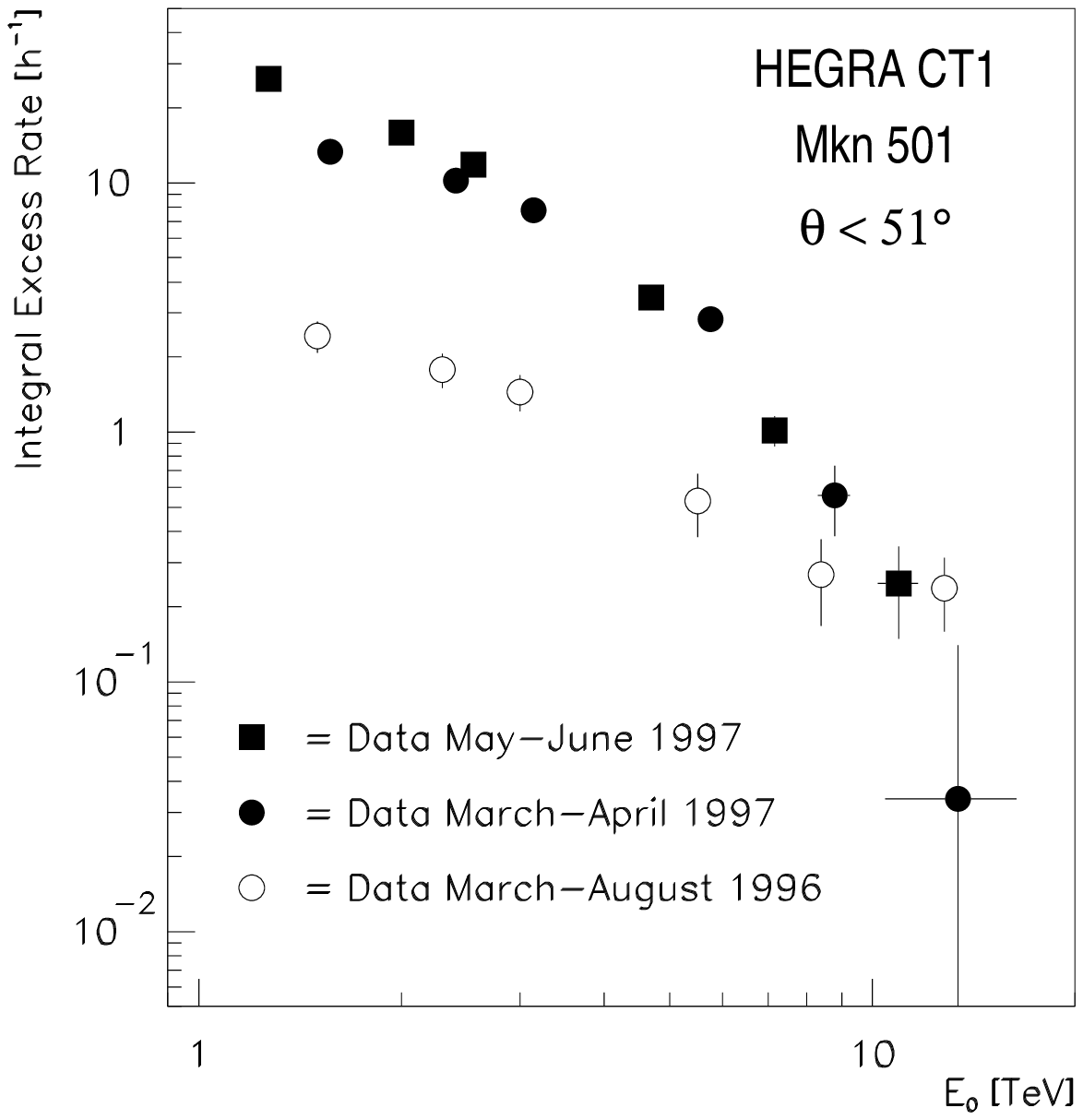}
\includegraphics{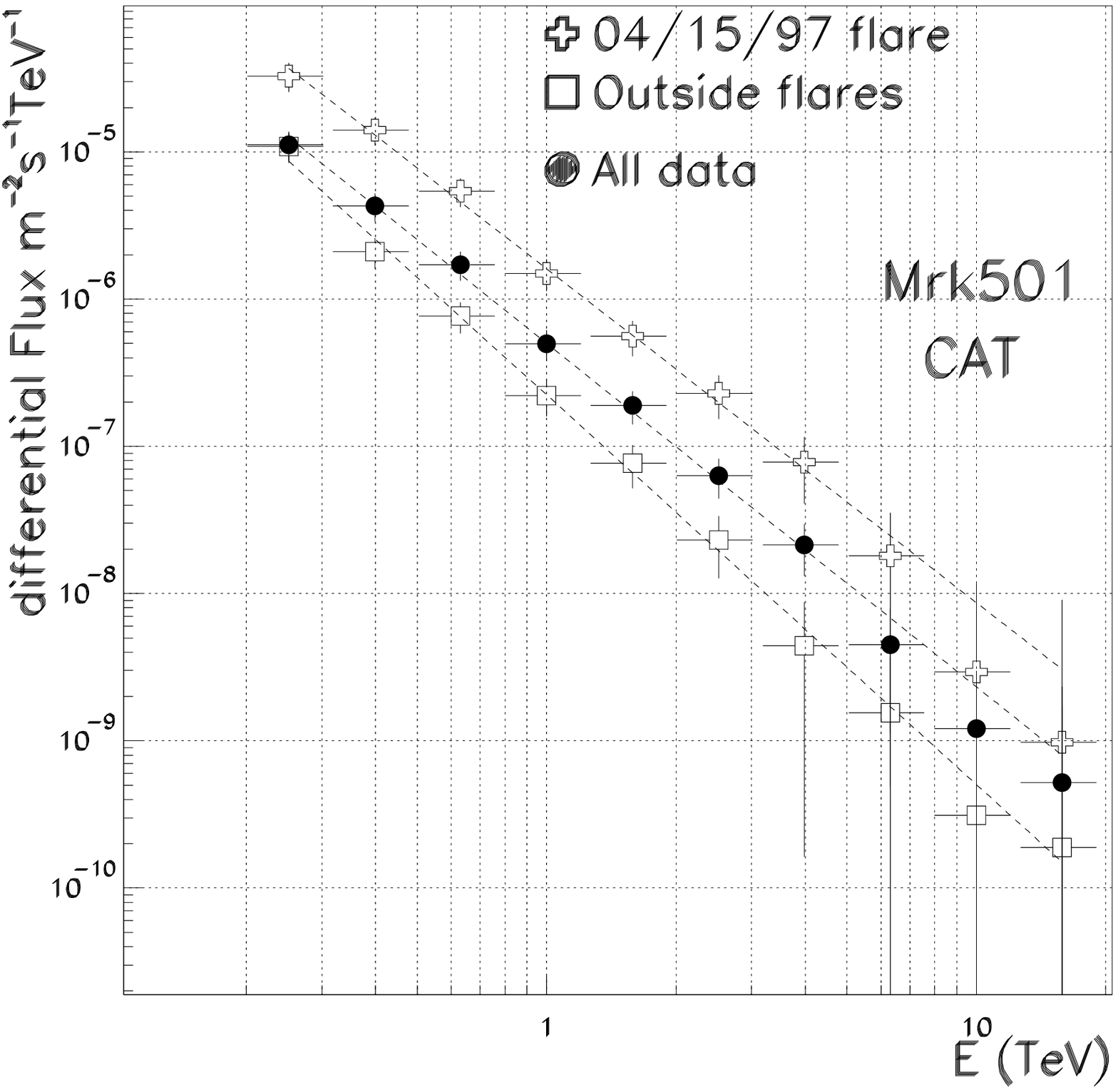}
\includegraphics{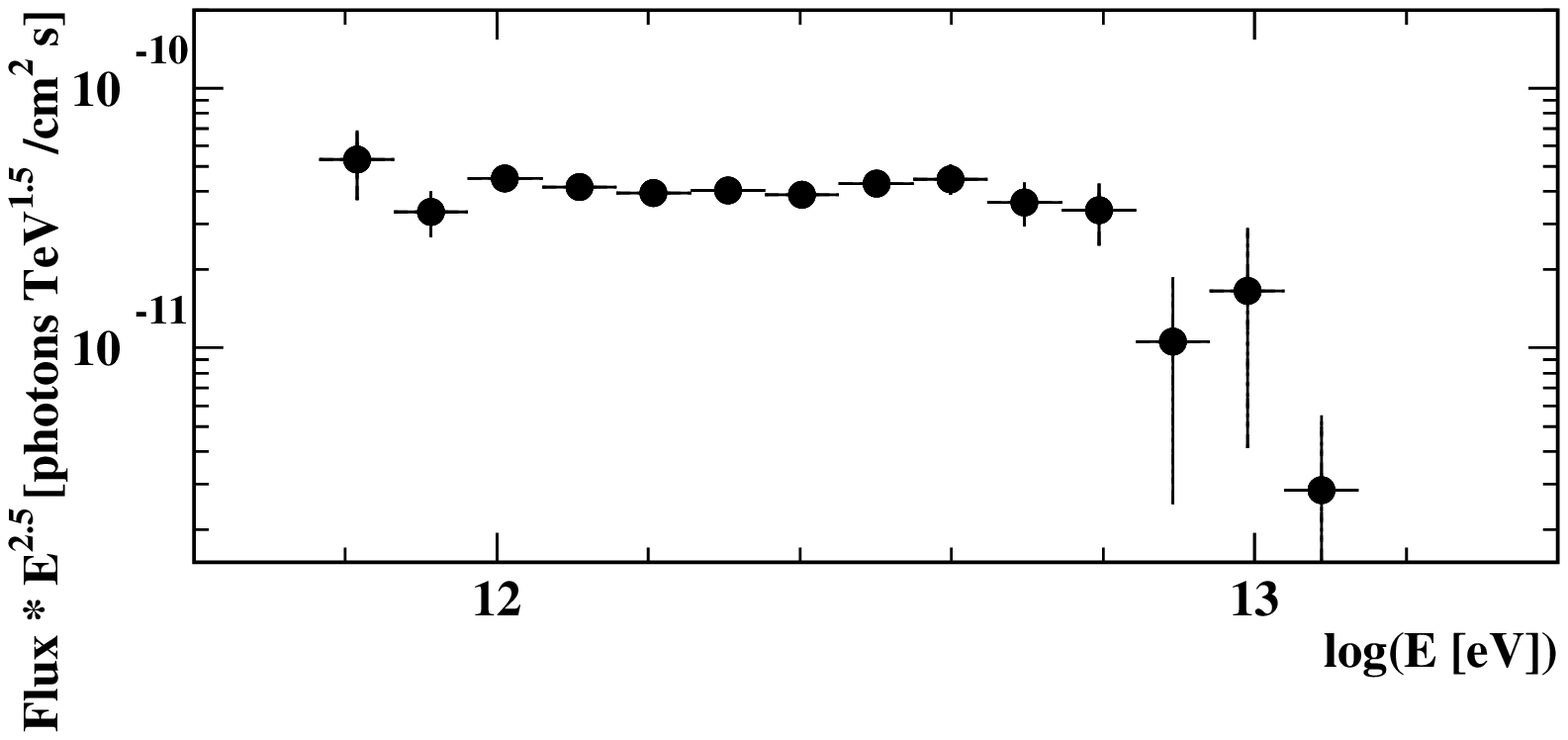}
\caption{(a) Top panel: preliminary integral spectrum obtained by
HEGRA CT1 in 1997 compared to that obtained in 1996.  (b) Middle
panel: preliminary spectra from CAT for the entire 1997 data set,
for 1997 periods between flares, and for the April 15/16 flare.
(c) Bottom panel: preliminary differential spectrum (multiplied
by $E^{2.5}$) obtained by the Telescope Array in 1997. }
\end{figure}

The differential energy spectrum obtained by the TA is shown in
Figure 5c and can be expressed $dF/dE = (4.0 \pm 0.2) \times
10^{-7} (E/1 \, {\rm TeV})^{-2.5 \pm 0.1}$ m$^{-2}$ s$^{-1}$
TeV$^{-1}$.  The spectrum becomes steeper above 5 TeV which may
suggest a cut off of the energy spectrum, although it is possible
that statistical fluctuations may cause this effect. More
statistics are needed to obtain conclusive results. 

The estimated time-averaged flux obtained from TACTIC
observations (April 9 - May 30, 1997) above a gamma-ray threshold
energy of $\sim (0.7 \pm 0.2)$ TeV  is nearly 2 Crab units, in
excellent agreement with the source spectrum inferred for the
corresponding period from independent observations by the HEGRA,
CAT and the Whipple groups.\\

\section{MULTI-WAVELENGTH OBSERVATIONS}
As a result of the high activity on Markarian 501 in March, 1997
a Target of Opportunity was declared for CGRO to observe the
source with all instruments for the period April 9-16. As
observed by several telescopes, Markarian 501 was very active
during this period in TeV gamma-rays with peaks on April 13 and
16; the flux level on April 16 was $> 10$ photons per minute.
The TeV light curve observed by Whipple, CAT and HEGRA is shown
in Figure 6a.  Note that HEGRA missed the flare on April 16 due
to cloudy sky.

\begin{figure}[htbp]
\vspace{20cm}
\includegraphics{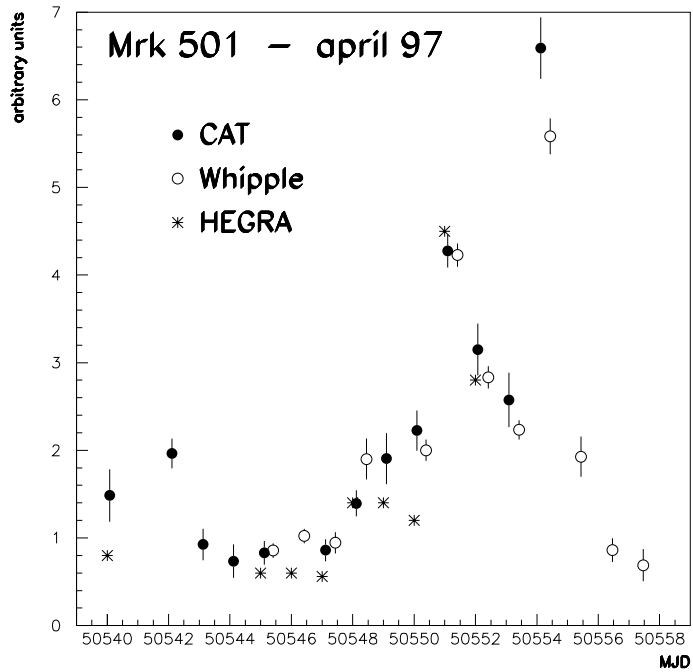}
\includegraphics{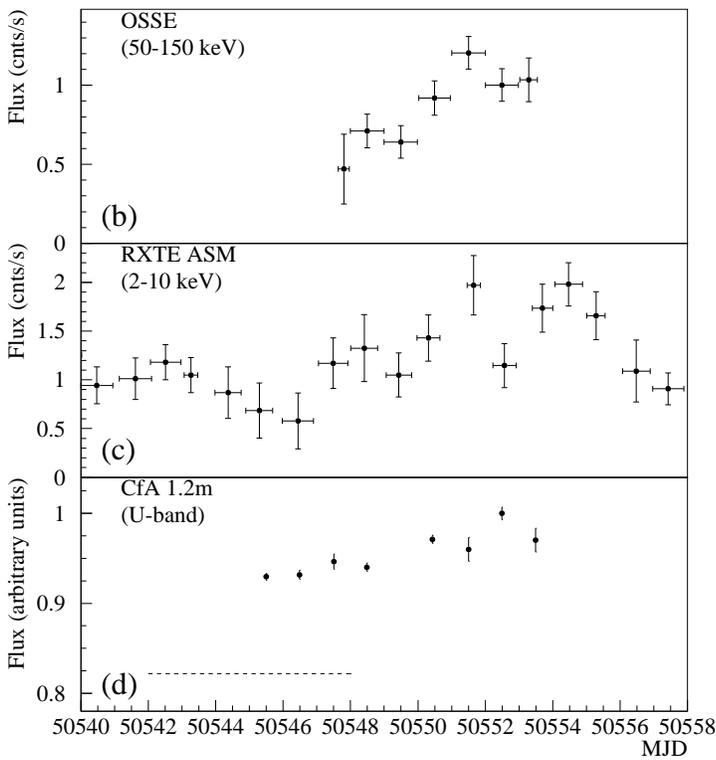}
\caption{(a) Top panel: VHE gamma-ray (Whipple, CAT, HEGRA), (b) OSSE
50--150 keV, (c) ASM 2--10 keV, and (d) U-band optical light
curves of Markarian 501 for the period 1997 April 2 (MJD 50540)
to 1997 April 20 (MJD 50558).  Dashed line in (d) indicates
average flux in 1997 March.  (b--d from Catanese et al. 1997.) }
\end{figure}

An analysis of the EGRET observations shows no statistically
significant evidence for emission, and an upper limit of $I(>100
\, {\rm MeV}) < 3.6 \times 10^{-7}$ photons cm$^{-2}$ s$^{-1}$
was derived. There was no report from COMPTEL, but OSSE, operating
in the 0.05-10 MeV range, saw a strong signal (Fig.~6b). The daily flux
varied by a factor of 2 (compared with the factor of 4 at TeV
energies). In the 50-150 KeV range OSSE saw the strongest signal
seen from any blazar. The spectral index is steep compared to
other blazars observed by OSSE (Figure 7).

\begin{figure}[htbp]
\vspace{12cm}
\includegraphics{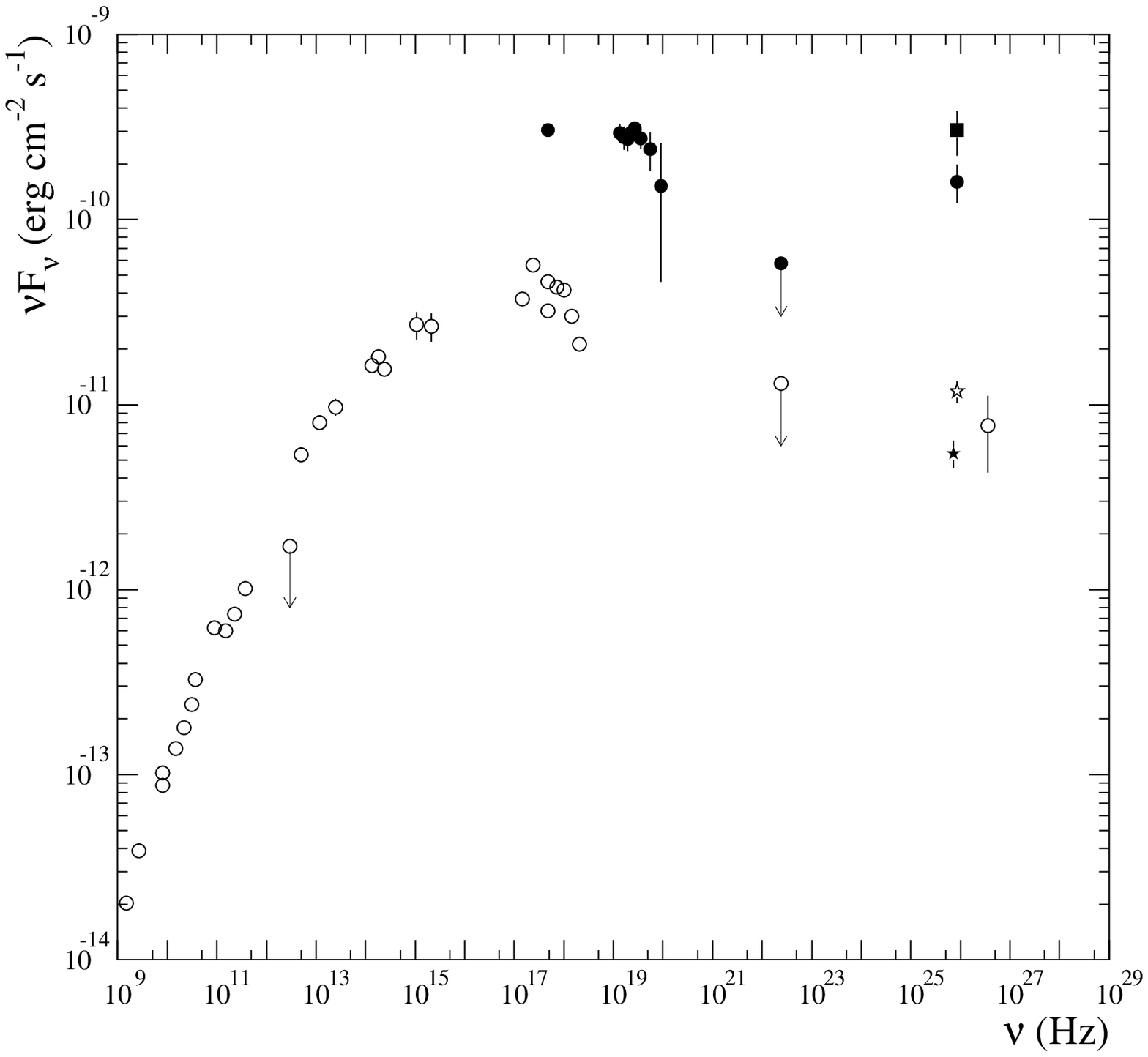}
\includegraphics{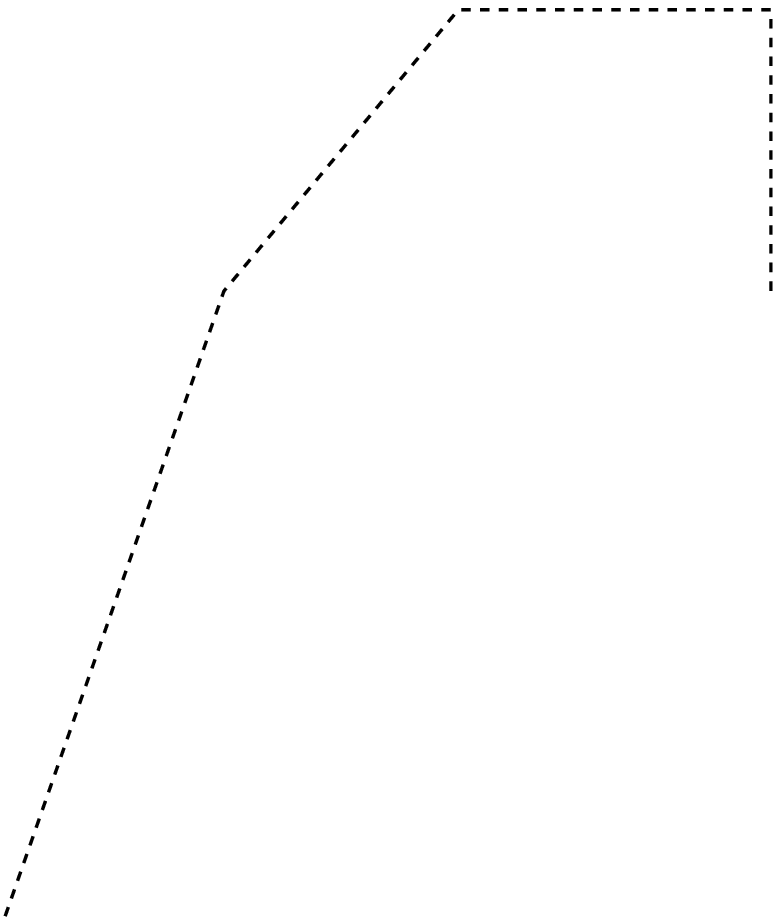}
\caption{The spectral energy distribution of Markarian 501:
contemporaneous observations taken April 9--15 (filled circels),
archival data (open circles), mean VHE gamma-ray flux in 1995
(filled star), mean flux in 1996 (open star), and maximum flux
detrected on April 16 (filled square).  (See Catanese et al for
references to data.)  The dotted line is the soft photon spectrum
assumed for April 9--15 in calculating photon-photon absorption.}
\end{figure}

The source was also observed by the RXTE (Catanese 1997) and Beppo
SAX (Pian et al. 1997) X-ray telescopes and the results will be
presented elsewhere. Results from the ASM detector on RXTE are
shown in Fig.~6c. There is a clear correlation between the
variations seen at TeV energies and those seen in X-rays; however
the amplitude of the variations is smaller at the longer
wavelengths. There is also weak evidence for correlation with
variations in the optical U-band (Fig.~6d).\\

\section{INTERPRETATION AND CONCLUSION}
The composite spectrum is shown in Figure 7. The double peak (at
UV-X-ray and gamma-ray energies) is characteristic of gamma-ray
emitting blazars; however the longer wavelength peak is shifted to
higher energies compared with Markarian 421 and other blazars.
Compared to the 1 keV cut-off in Markarian 421 the Markarian 501
cut-off is $> 100$ keV, implying in a Compton-synchrotron model that
the electrons have very high energies. In this case, the lack
of detection of 100~MeV gamma rays by EGRET results from this energy
falling between the synchrotron and inverse Compton peaks.

The very rapid time-variability observed places strong
constraints on models of gamma-ray emission.  The fact that only
blazars appear to show strong gamma ray emission, and that this
class of AGN has relativistic jets closely aligned with the line
of sight, strongly suggests the gamma rays originate in an
emission region in the jet.  Because of the lack of emission
lines it is likely that much of the lower energy radiation is
also non-thermal.  In the homogeneous synchrotron-self Compton
model the low energy photons are produced in the same region as
the high energy emission.  We shall consider the emission region
to be a ``blob'' or radius $r$ moving relativistically along the
jet axis with velocity $\beta_b c$ and Lorentz factor
$\Gamma_b=(1 - \beta_b^2)^{-1/2}$, and containing a population of
relativistic electrons, magnetic field and radiation.

If an AGN at redshift $z$ is observed at angle $\theta$ to the
jet axis and the blob emits two pulses of light separated by
blob-frame time $\Delta t'$ (blob frame quantities are primed),
the observed arrival times of the two pulses will be separated
by $\Delta t_{\rm obs} = D^{-1} \Delta t'$ where
\begin{equation}
D = [(1+z) \Gamma_b (1 - \beta_b \cos \theta)]^{-1}
\end{equation}
is the ``Doppler factor''.  The energies of emitted photons are 
also boosted in energy as a result of the bulk motion of the 
blob such that $\varepsilon = D \varepsilon'$.

If $\Delta t_{\rm obs}$ is the fastest observed variability time,
and we assume the emissivity varies with time uniformly over the
blob, then the radius of the blob can not be much larger than $r
\approx 0.5 c D \Delta t_{\rm obs}$.  Then the differential
photon number density at energy $\varepsilon'$ in the blob frame per
unit photon energy is approximately
\begin{equation}
n(\varepsilon') \approx {4 d_L^2 F(\varepsilon) 
\over c^3 \Delta t_{\rm obs}^2 D^4}
\end{equation}
where $F(\varepsilon)$ is the observed differential photon flux at
energy $\varepsilon = D \varepsilon'$ and $d_L$ is the luminosity
distance.  If the Doppler factor is too small then the photon
density in the blob frame may be so large that gamma-rays may not
escape due to photon-photon collisions.  The optical depth for this
process depends on energy $E$ and is given by
\begin{equation}
\tau_{\gamma \gamma}(E) \approx {r \over 8 {E'}^2} 
\int d \varepsilon' {n(\varepsilon') \over {\varepsilon'}^2}
\int ds s \sigma_{\gamma \gamma}(s)
\end{equation}
where $E = DE'$, and $\sigma_{\gamma \gamma}(s)$ is the cross
section for photon-photon pair production at centre of momentum
frame energy squared $s$.  During April the 2--10 keV and 15--150
keV fluxes were at roughly the same level with the 2--10 keV flux
being about 5 times higher than the archival data.  Assuming the
lower energy fluxes increased by about the same amount, one
obtains the spectral energy distribution given by the dotted
curve which has been added to Figure 7. One can then calculate
the minimum Doppler factor as a function of gamma-ray energy,
such that $\tau_{\gamma \gamma}(E)=1$, for any given variability
time scale.  This has been done for $\Delta t_{\rm obs}=1$ day, 1
hour, and 15 minutes (Bednarek and Protheroe 1998) and is shown
in Figure 8.  We note that for a 15 minute variability time scale
and a gamma ray spectrum extending to 20 TeV that a Doppler
factor of at least about 30 is required.

\begin{figure}[htbp]
\vspace{8cm}  
\includegraphics{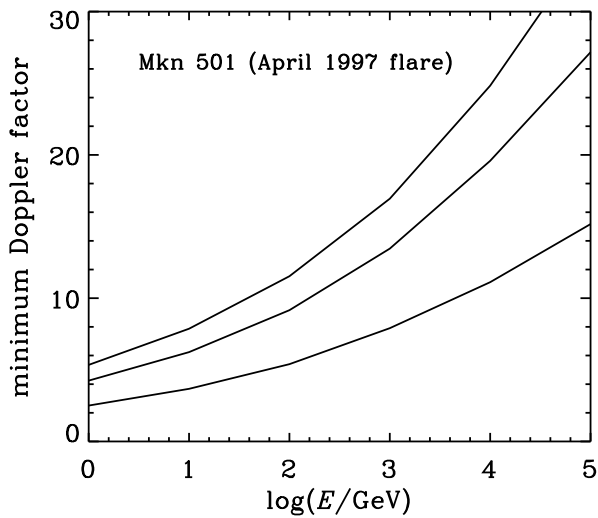}
\caption{Minumum Doppler factor versus gamma-ray energy for
$\Delta t_{\rm obs}=1$ day (bottom curve), 1 hour, and 15 minutes
(top curve). }
\end{figure}

Gamma ray emission from active galactic nuclei (AGN) is often
interpreted in terms of the homogeneous synchrotron
self-Compton model (SSC) in which the low energy emission (from
radio to X-rays) is synchrotron radiation produced by electrons
which also up-scatter these low energy photons into high energy
$\gamma$-rays by inverse Compton scattering (IC) (Macomb et
al.~1995, Inoue \& Takahara~1996, Bloom \& Marscher~1996,
Mastichiadis \& Kirk~1997). In this model all the radiation comes
from this same region in the jet. Such a picture can naturally
explain synchronized variability at different photon energies and
has been tested in the context of Markarian 421 (Bednarek and Protheroe
1997a). More complicated (inhomogeneous) SSC models are also
proposed which postulate that the radiation at different energies
is produced in different regions of the jet (e.g.  Ghisellini et
al.~1985, Maraschi et al.~1992).

In the SSC model the spectral energy distribution has two broad
peaks.  The low energy component is due to synchrotron emission,
and the high energy component being due to inverse Compton
scattering of the low energy component by the same electrons that
produce the synchrotron radiation.  We identify the low energy
component with the photon spectrum extending up to $\varepsilon_{\rm
max} \sim 0.2$~MeV observed by OSSE, and the high energy
component with the TeV gamma rays.  The inverse Compton
scattering producing the highest energy gamma rays at $E_{\rm
max} \sim 20$~TeV will be in the Klein-Nishina regime where the
scattered photon energy is comparable to the maximum electron
energy.  Thus
\begin{equation}
E_{\rm max} \sim D {\gamma'}_{\rm max} mc^2
\end{equation}
where ${\gamma'}_{\rm max} mc^2$ is the maximum electron energy
in the jet frame.  Thus we obtain ${\gamma'}_{\rm max} \sim 1.3
\times 10^6$ for $D = 30$ and $E_{\rm max} \sim 20$~TeV.

The maximum energy of the low energy component will be determined by
the magnetic field in the blob $B'$ and by ${\gamma'}_{\rm max}$ and is 
given approximately by
\begin{equation}
\varepsilon_{\rm max} \approx 0.5 D {\gamma'}_{\rm max}^2 \varepsilon'_B
\end{equation}
where $\varepsilon'_B = (B' / B_{\rm cr}) mc^2$, and $B_{\rm
cr}=4.414 \times 10^{13}$~G.  Thus we can obtain the magnetic field
in the blob,
\begin{equation}
B' \approx {2 \over D {\gamma'}_{\rm max}^2} {\varepsilon_{\rm max}
\over mc^2} B_{\rm cr},
\end{equation}
which gives $B' \approx 0.7$~G for $D = 30$, ${\gamma'}_{\rm max}
= 1.3 \times 10^6$ and $\varepsilon_{\rm max} = 0.2$~MeV.

To obtain the observed variability time (in particular the
observed fall time of the gamma ray signal) the highest energy
electrons must be able to convert their energy into radiation in
a time less than that observed.  Thus we require at least that
\begin{equation}
t'_{\rm cool} < D \Delta t_{\rm obs}.
\end{equation}
We now have an estimate of the differential photon
number density and the magnetic field in the blob, and are able
to calculate the cooling times of electrons by inverse Compton
scattering and synchrotron radiation.  The cooling time in the
blob frame for each process is given by
\begin{equation}
t'_{\rm cool} \approx {\gamma' mc^2 \over 1.33 \sigma_T U' {\gamma'}^2}
\end{equation}
where $U'$ is either $U'_{\rm mag} = {B'}^2/8\pi \approx 0.017$
erg cm$^{-3}$ which is the magnetic energy density (synchrotron losses)
or
\begin{equation}
U'_{\rm rad}(< \varepsilon'_T) \approx \int_0^{\varepsilon'_T} n(\varepsilon')
\varepsilon' d \varepsilon' \approx 0.033 \; \; {\rm erg \,cm^{-3}}
\end{equation}
which is the energy density of photons which will scatter in the
Thomson regime, where $\varepsilon'_T = mc^2/\gamma'$.  Thus we obtain
fall times in the observer's frame of $t'_{\rm cool}/D \approx
25$~s and 50~s for inverse Compton and synchrotron respectively,
which are consistent with the observed variability time scale of
several minutes.

We should also check whether electrons can be accelerated up to
the maximum energy given the magnetic field present in the blob.
This depends on their acceleration rate
\begin{equation}
\dot{E}'_{\rm acc} = \chi e c B',
\end{equation}
where $\chi$ is the acceleration efficiency which depends on the
mechanism, and on the synchrotron plus IC energy loss rate 
\begin{equation}
\dot{E}'_{\rm loss} \approx - 1.33 \sigma_T c \alpha U'_{\rm mag} {\gamma'}^2
\end{equation}
where $\alpha = [1 + U'_{\rm rad}(< \varepsilon_T)/U'_{\rm mag}] \approx 2.9$.
Equating $\dot{E}'_{\rm acc}$ to $-\dot{E}'_{\rm loss}$ we obtain
\begin{equation}
\gamma'_{\rm max} \approx 1.5 \times 10^8 (\chi/\alpha)^{(1/2)} {B'}^{(-1/2)}.
\end{equation}
For $\gamma'_{\rm max} > 1.3 \times 10^6$ we require $\chi/\alpha
> 5 \times 10^{-5}$.  Such an acceleration efficiency is quite reasonable
for shock acceleration.

The acceleration time should be consistent with the rise time of
the the observed gamma ray signals, i.e. $t'_{\rm acc} =
E'/\dot{E}' < D \Delta t_{\rm obs}$.  The acceleration time is given by
\begin{equation}
t'_{\rm acc} = 5.7 \times 10^{-8} (\chi B')^{-1} \gamma' \; \; {\rm s}
\end{equation}
and for $\gamma' =1.3 \times 10^6$, $B' = 0.7$~G, $D=30$ and
$\chi = 10^{-4}$ we obtain $t'_{\rm acc}/D \approx 35$~s which is
again consistent with the observed rise time.

From the discussion above one would conclude that the SSC model
can give a satisfactory explanation for the present observations
of Markarian 501.  However, the model parameters are rather
finely balanced.  The comparable cooling times for inverse
Compton and synchrotron radiation lead to similar power being
emitted at TeV energies and sub-MeV energies as was observed
during April 1997 (see Fig.~7).  If, however, more rapid
variability were observed or much higher energy gamma rays were
detected on the same variability time scale, then the minimum
Doppler factor required for gamma ray escape would increase (see
Fig.~8) such that the SSC model could be in trouble.  The reason
for this is that the balance between $t'_{\rm cool}$ for
synchrotron and IC (which are proportional to ${U'}_{\rm mag}^{-1}$
and ${U'}_{\rm rad}^{-1}$) is disturbed.  For example, from Eq.~5,
$\gamma'_{max} \sim D^{-1}$ and from Eq.~6 $B' \sim D^{-1}
{\gamma'}^{-2}_{max}$ giving $U'_{\rm mag} \sim D^2$, while
$U'_{\rm rad} \sim D^{-4}$.  Thus the ratio of the power going
into the high energy component to that going into the low energy
component is proportional to $D^{-6}$.  Increasing $D$ by 50\%
causes this ratio go down by a factor of 10.  One should also not
overlook other possible scenarios such as hadronic models
(Mannheim 1995, Protheroe 1996) which can be tested by searching
for correlated high energy neutrino emission.

If the 13 day or 29 day periodicity suggested by the analysis of
the TA and HEGRA light curves is confirmed it will provide important
constraints on the emission region/mechanism.  One possibility
under consideration (Protheroe 1998) is that this may
arise if the jet has internal helical structure, or is helical as
has been suggested by Conway and Wrobel (1995) based on the
observed mis-alignment of the pc scale and kpc scale jets in
Markarian~501.  However, for this to work, i.e. to give the
periodicity and for the jet to have the observed jet
mis-alignment, one would require a helix wavelength of about 50
kpc, a viewing angle of $\sim 10^{-3}$, and a jet Lorentz
factor of $\sim 10^3$.

Finally, we mention some alternative scenarios which might
possibly explain the rapid variability observed in both Markarian 421
and Markarian 501.  Bednarek and Protheroe (1997b,1997c) have suggested
that this could arise as a result of the interaction of stars
with the jet, or as a result of time dependent gamma ray
absorption by collisions of the gamma rays with photons from from
a hot-spot on an accretion disk.\\

\section{REFERENCES}
Bednarek, W., and Protheroe, R.J., (1997a) MNRAS, in press \\
Bednarek, W., and Protheroe, R.J., (1997b) MNRAS, 287, L9\\ 
Bednarek, W., and Protheroe, R.J., (1997c) MNRAS, 290, 139\\ 
Bednarek, W., and Protheroe, R.J., (1998) MNRAS, in preparation\\ 
Bloom, S.D., Marscher, A.P., (1996) ApJ 461, 657\\
Bradbury, S.M. et al. (1997) A\&A, 320, L5. \\
Breslin, A.C. et al. (1997) IAU Circ. 6596. \\
Carter-Lewis, D.A. et al., (1997) 25th ICRC (Durban), 3, 161. \\
Catanese, M. (1997) in preparation. \\
Catanese, M., et al. (1997) ApJL (in press). \\
Conway, J.E., and Wrobel, J.M., (1995) Ap.J., 439, 112\\
Ghisellini, G., Maraschi, L., Treves, A., (1985) A\&A, 146, 204\\
Inoue, S., Takahara, F., (1996) ApJ 463, 555\\
Macomb, D.J. et al., (1995) ApJ, 449, L99 (Erratum 1996, ApJ 459, L111)\\
Mannheim, K, (1995) Astroparticle Phys., 3, 295\\
Maraschi, L., Ghisellini, G., Celloti, A., (1992) ApJ 397, L5\\
McEnery, J. et al. (1997) 25th ICRC (Durban) 3, 257. \\
Mastichiadis, A., Kirk, J.G., (1997) A\&A, in press\\
Pian, E., et al. (1997) ApJL, submitted. \\
Protheroe, R.J.,  (1996) in Proc. IAU Colloq. 163, Accretion
Phenomena and Related Outflows,\\ 
\indent ed. D. Wickramasinghe et al., in press\\
Protheroe, R.J. (1998) in preparation.\\
Quinn, J. et al. (1996) ApJ, 456, L83. \\
Quinn, J. et al. (1997) 25th ICRC, 3, 249. \\
Quinn, J. (1997) National University of Ireland, Ph.D. Dissertation
(unpublished). \\

\end{document}